\documentclass[sigconf]{acmart}

\renewcommand\footnotetextcopyrightpermission[1]{} 
\setcopyright{none}

\usepackage{microtype}
\usepackage{enumitem}
\usepackage{booktabs} 
\usepackage{hyperref}
\usepackage{flushend}
\usepackage{subfig}
\usepackage{algorithm}
\usepackage{algorithmic}
\usepackage{multirow}
\usepackage{colortbl}
\usepackage[most]{tcolorbox}
\usepackage{dblfloatfix}
\usepackage{pifont}
\usepackage{tikz}
\usepackage{pgfplots}
\usepackage{enumitem}
\usepackage{wasysym}
\usepackage{xspace}
\usetikzlibrary{backgrounds, positioning, fit}
\usetikzlibrary{shapes.geometric}
\usepackage{amsmath}
\usetikzlibrary{patterns}
\usetikzlibrary{pgfplots.groupplots}

\newif\ifcomment
\commenttrue
\ifcomment
\newcommand{\virat}[1]{{\color{red} \textbf{Virat: #1}}}

\else
\newcommand{\virat}[1]{}
\fi

\begin{document}

\title{Privacy Assessment of Federated Learning using Private Personalized Layers}

\author{Th\'eo Jourdan}
\affiliation{%
  \institution{Univ Lyon, INSA Lyon, Inria, CITI, CREATIS, Inserm}
}
\email{theo.jourdan@insa-lyon.fr}

\author{Antoine Boutet}
\affiliation{%
	\institution{Univ Lyon, INSA Lyon, Inria, CITI}
}
\email{antoine.boutet@insa-lyon.fr}

\author{Carole Frindel}
\affiliation{%
	\institution{Univ Lyon, INSA Lyon, CREATIS, Inserm}
}
\email{carole.frindel@creatis.insa-lyon.fr}

\begin{abstract}
Federated Learning (FL) is a collaborative scheme to train a learning model across multiple participants without sharing data.
While FL is a clear step forward towards enforcing users’ privacy, different inference attacks have been developed.
In this paper, we quantify the utility and privacy trade-off of a FL scheme using private personalized layers.
While this scheme has been proposed as local adaptation to improve the accuracy of the model through local personalization, it has also the advantage to minimize the information about the model exchanged with the server. However, the privacy of such a scheme has never been quantified.
Our evaluations using motion sensor dataset show that personalized layers speed up the convergence of the model and slightly improve the accuracy for all users compared to a standard FL scheme while better preventing both attribute and membership inferences compared to a FL scheme using local differential privacy.
\end{abstract}

\keywords{Privacy, Federated Learning, Inference Attacks}

\maketitle
\pagestyle{plain}



\section{Introduction}
\label{sec:intro}

The development of Internet of Things (IoT) and wearables contributed to the progress of a wide range of quantified-self applications related to activity recognition. 
In this context, motion data extracted from connected devices (e.g., smartwatches) equipped with motion sensors (e.g., accelerometer and gyroscope) are sent to a central server which processes these data through machine learning models. According to the considered application, these learning models can classify which activity is performed by the user or compute other information such as the number of steps or burned calories.

While quantitative analysis of daily activities can bring benefits from the health perspectives~\cite{health1,health2}, transferring all this data to a third-party server raises important privacy concerns. 
Indeed, data breaches, compromised servers or any unwanted exploitation of the data expose users to personal and sensitive information leakage such as health-related attributes~\cite{privacyhealth}. 

To mitigate this risk, a Federated Learning (FL) architecture (also known as collaborative learning) has been proposed~\cite{mcmahan}.
In this scheme, the personal data of the user stays locally on its device and only a learning model is exchanged with the server.
Iterativelly, the server sends a model to devices, this model is trained and refined with the local data on each device. Model updates are sent to the server which aggregates them to maintain a global learning model which will be disseminates back a devices.
While this iterative process works well in case of data sharing similar distribution, heterogeneity of data across user devices can severely degrade performance of standard federated averaging for ML learning applications, especially for atypical users. 
Indeed, one unique model cannot cope with the heterogeneity of data and provide the best utility for all users~\cite{dysan}. 
To address this data heterogeneity, several local adaptation schemes have been proposed such as fine-tuning of personalized layers, multi-task learning, and knowledge distillation~\cite{fedper,impact2} which depict a benefit for all participants in terms of accuracy.

While FL improves privacy by reducing the exposition of the personal data, it remains vulnerable to threats.
For instance, FL is not robust to model poisoning which aims to destroy the convergence of the central model~\cite{blanchard,bernstein}. 
Privacy leakages may also occur through membership inference attacks~\cite{membership} which consist of inferring the presence of an individual data record in the training data, or attribute inference attack where the adversary is able to infer sensitive information about individuals~\cite{attribute}. 
The adversary can be passive~\cite{passiveactive1} (i.e., only observing exchanges) or active~\cite{passiveactive2} (i.e., modifying the protocol), and can control users or the server.
To mitigate the risks, several approaches have been proposed from using Differential Privacy locally at user level or server level~\cite{naseri}, Homomorphic Encryption (HE) and Secure Multiparty Computation (SMC) \cite{hc_smc}.



In this paper, we quantify the utility and privacy of a FL scheme using private personalized layers~\cite{fedper}.
In such a scheme, only the lower layers of the model (capturing coarse grain information) are exchanged with the server while the upper layers of the model (capturing fine grain information) are personalized and kept private on each user.
This scheme is known to improve the accuracy of the model in presence of heterogeneous data across users.
However, the privacy impact of sharing only a sub part of the model has never been measured.
To assess privacy leakage of this scheme, we consider both an attribute and a membership inference attack.
Evaluations have been conducted using two datasets of motion sensor data collecting in real-life conditions.
Results show that FL with personalized layers speeds up the convergence compared to vanilla FL and slightly increases the activity accuracy between 1\% and 5\%, while decreasing the gender and the overweight inference between 10\% and 20\% and 15\% on average for membership inference.
This utility and privacy trade-off is better than a defense scheme using local differential privacy which decreases the inference of the gender and the overweight up to 12\% but at the cost of the activity accuracy which reduces up to 10\%.
These results tend to show that minimizing the information exchanged with the server is an interesting avenue for both personalizing the model (i.e., improving accuracy) while limiting potential inferences (i.e., improving privacy).


The outline of the paper is as follows. First, background and related work are described in Section~\ref{sec:background}. The exhaustive evaluation is then presented in Section~\ref{sec-eval} before concluding in Section~\ref{ending}.

\section{Background and Related Work}
\label{sec:background}

In this section, we review background and related work on FL (Section~\ref{sec-fl}) and inference attacks and mitigation schemes (Section~\ref{sec-inference-defense}).

\subsection{Federated Learning}
\label{sec-fl}

Deep Neural Networks are now the most effective algorithms for a lot of machine learning tasks. 
A new paradigm has been proposed by training Machine Learning (ML) models on the user devices, named Federated Learning (FL). 

In a FL scheme following~\cite{mcmahan}, participants keep their data locally on their device and exchange a model -- targeting a specific learning task -- with a server.
The main objective of this algorithm is to iteratively train a learning model \textit{M} maintained by the server by aggregating this model trained locally on each participant.
At each learning round \textit{i}, each client \textit{k} trains its local model \textit{$m_k$} with its own data using Stochastic Gradient Descent (SGD) during several iterations \textit{j}. 
In its synchronous version, once all the participants send their model update \textit{m} to the server, this server then aggregates all these model updates using the following equations before to disseminate back this aggregated model to all devices:

$$ M_{i+1} = \sum_{c=1}^{C} \frac{n_c}{n} m_c^{i+1},$$

with $n_c$ the set of indexes of all the data points $n$ on client $c$, $m_c^{i+1}$ the local update of a client $c$, calculated with the following equation:

$$ m_c^{i+1} = m_c^{i} - \eta g_c^i ,$$



with $\eta$ a fixed learning rate (i.e., hyperparameter which controls the step size of the optimization) for each client and $g_c^i$ the average gradient on the local data of the client $c$ at the epoch $i$.
Those learning rounds continue until the convergence of the central model.


\begin{figure}[!h]
\centering
\includegraphics[width=6.5cm]{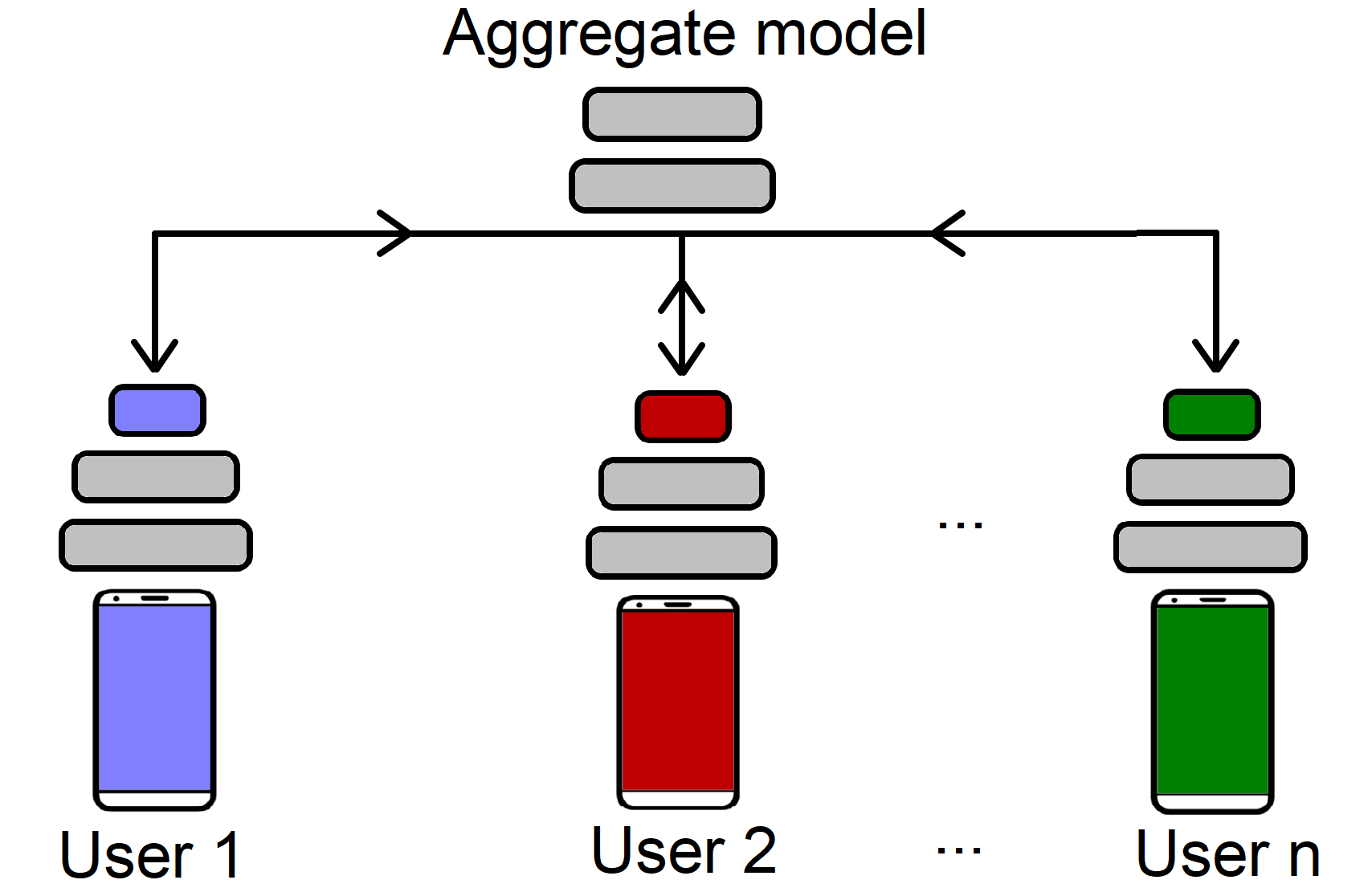}
\caption{Personalized FL approach: only the upper layers (colored in grey) are shared with the server while the personalization layers are kept private on the device.}
\label{fig:fl_flow}
\end{figure}

The development of FL highlighted other challenges~\cite{review_fl} such as the heterogeneity of data across user devices leading to a degraded accuracy for less represented users.
To overcome this limitation, \cite{fedper} studied an FL scheme using personalization layers.
In this scheme, the local model on each participant is composed of \textit{lower layers} trained following classical FL learning round, and \textit{upper and personalization} layers trained locally and which stay private on the device and not exchanged with the server (Figure~\ref{fig:fl_flow}). 


\subsection{Inference Attacks and Defenses}
\label{sec-inference-defense}

Privacy concerns are another main challenge of FL. 
Even though the raw data is not shared but kept on each local device, the model updates exchanged between participants and the server can leak sensitive information. 
Specifically, two types of attack can be considered. 
Poisoning attacks from malicious users aim at preventing the convergence of the learning model~\cite{blanchard}, or implanting a backdoor to control its behaviour~\cite{backdoor}. 
Inference attacks, in turn, attempt to infer sensitive information on the users through model updates exchanged during the training process. These attacks can be conducted by participants or by the server. In addition, they can be passive or active.
For instance, a malicious server can infer sensitive information based on the local updated parameters $w^i$ sent by each user. To increase the observation, the server can send to users a fudged model in order to amplify the potential inference of sensitive information ~\cite{passiveactive2}.
There are mainly two types of inference attacks: the attribute and the membership inference attacks.
Attribute inference attack consists of inferring a sensitive information of the user~\cite{attribute} while membership inference attack consists of determining if a data record has been used for the training of a specific 
model~\cite{membership}.
To apply these attacks on a FL scheme, a supervised classification model (i.e., Random Forest) was trained to infer the sensitive information from the model's parameters shared as input.

To prevent these attacks, several approaches have been considered. 
Homomorphic Encryption (HE) and Secure Multiparty Computation (SMC)~\cite{hc_smc} approaches have been adapted to FL but are difficult to apply at scale due to overhead. 
Local Differential Privacy (LDP)~\cite{naseri}, in turn, 
consists of introducing a random perturbation on SGD during the learning phase on the user device. 
The method provides statistical guarantees on the inference capability of an adversary to infer information on an individual's data. 
A differential-private SGD can be written as:

$$ w_{t+1} = w_t - \eta _t (\nabla l(w_t, x_t, y_t) + N_t) ,$$

with $\eta_t$ the learning rate, $w$ the parameter that minimizes the objective function $l$ for a data point ($x$,$y$) at time $t$, and $N_t$ a noise value that follows a Gaussian distribution.

\section{Evaluation}
\label{sec-eval}

We exhaustively evaluate the utility and privacy of a FL scheme using private personalized layers in the context of activity recognition (details of the methodology are given Section~\ref{settings}).
In this section, we show that personalized layers improves the utility (Section~\ref{utility}) and privacy (evaluated through attribute inference Section~\ref{privacy-attribute} and membership inference Section~\ref{privacy-mia}) compared to both a vanilla FL and a defense scheme using local differential privacy.

\subsection{Experimental setting}
\label{settings}

\textbf{System model:} We consider a FL scheme using Stochastic Gradient Descent (SGD) addressing activity recognition.
The learning model is based on 2 convolutional layers, and 3 fully connected layers. 
Only the 2 lower convolutional layers are exchanged with the server which aggregates and disseminates model updates to devices, the 2 upper fully connected layers stay private on the device and are personalized with the user data. 
The devices of users are considered as trusted but it is not the case of the server which is considered as an adversary trying to infer personal information of participants from their model updates.


\setlist{nolistsep}
\textbf{Datasets:}
Two real-life condition datasets are used for the evaluation. They are both publicly available and heavily used in the literature. These datasets come from the extraction of motion sensor data during gait activities (\textit{i.e.}, based on step patterns) of different subjects.

\begin{itemize}
\item \textbf{MotionSense}~\cite{motionsense} contains motion data captured from an accelerometer (\textit{i.e.}, acceleration and gravity) and gyroscope of an iPhone 6s kept in the front pocket at a frequency rate of 50Hz.  
Overall, six activities (\textit{i.e.}, walking, jogging, going upstairs, going downstairs, sitting and standing) have been made by 24 users during 15 trials in the same conditions and environment.
\item \textbf{MobiAct}~\cite{mobiact} records the motion data from 58 subjects during more than 2500 trials, all captured with a smartphone also in the front pocket.
This dataset includes signals recorded from the accelerometer and gyroscope of a Samsung Galaxy S3 smartphone. Nine different activities of daily living are performed by the users. 
We only used the trials corresponding to the same activities as MotionSense in order to do the evaluation with the exact same settings. 
\end{itemize}

Both datasets contain an equal number of men and women, and each activity is performed according the same conditions by all subjects.
However, the walking activity is more represented than the others. For each user, we also have access to physical information (e.g., the gender, weight, height and age).  


\textbf{Baselines:}
We considered two baseline approaches to compare FL scheme using private personalized layers (\textbf{FedPer})~\cite{fedper}:

\begin{itemize}

\item \textbf{Standard FL (Vanilla)}~\cite{mcmahan} This is the most common FL scheme using SGD training on the device and average aggregation of all models at each learning round on the central server.

\item \textbf{Local Differential Privacy (LDP)}~\cite{naseri} We consider an implementation based on an introduction of noise following a Gaussian distribution ($\mathcal{N}$(0,0.01)) to the model updates computed through a classical learning phase (i.e., Vanilla).

\end{itemize}




\textbf{Evaluation metrics:}
We evaluated FedPer and the different baselines along both utility and privacy metrics.

\begin{itemize}
\item \textbf{Utility: } To measure the utility, we considered the accuracy of the predicted activity. 
More precisely, we produce a confusion matrix based on the output of the classifier and measure the number of correct predictions made by this classifier over all predictions made. The value of the accuracy ranges from $0$ to $1$, in which $1$ corresponds to perfect accuracy. 

\item \textbf{Privacy: } To assess the level of privacy, we rely on the accuracy of both the inference of sensitive attributes and the inference to be a member of the training set. These inference attacks implement the solution proposed by~\cite{attribute} which leverages an invariant permutation representation of nodes at each layer to classify model updates received by the server through a random forest of 1000 trees with a maximum depth of 10. 
We consider the gender and the Body Mass Index (BMI) of the users as sensitive attributes. The BMI is a value defined by the weight of the user divided by the square of her height. This value allows to categorize a person as underweight, normal weight, overweight or obese. In our case we only focus on a binary classification: overweight (BMI $>$ 25) or not (BMI $<$ 25) for the sake of class balance. 
For the membership inference, the accuracy refers to the percentage of correct prediction (that a participant has been involved in the training of the model) over all predictions made.
In both attacks, an accuracy of $0.5$ corresponds to a random guess as our dataset is balanced. 
\end{itemize}



\textbf{Implementation details:}
For each experiment, we run 10 times of 5-fold cross validation where each fold is tested based on the training of the other four.
We considered 200 learning rounds and an early stopping that stops the learning process if the average test loss of the aggregated model sent locally on the user data does not decrease during 30 learning rounds. 
During each learning round, the training with SGD is done locally at user's level during 10 epochs.
A constant learning rate is used with $\eta = 0.001$ for all the users. 
These parameters have been optimized independently for the tasks of activity recognition, gender and BMI inference.

\subsection{Utility Evaluation}
\label{utility}

\begin{figure}[!h]

\begin{minipage}[b]{0.43\linewidth}
  \centering
  \centerline{\includegraphics[width=4.7cm]{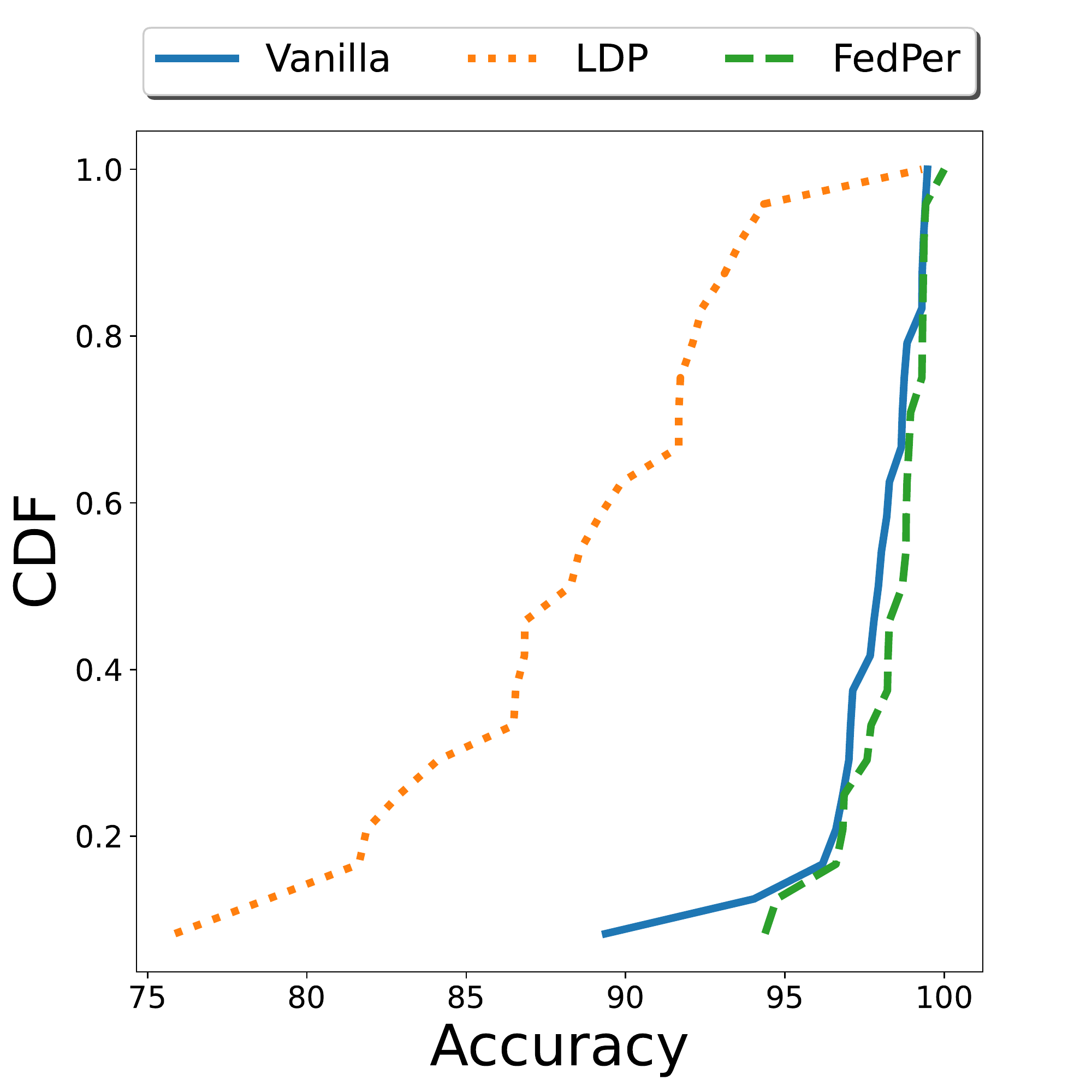}}
  \centerline{(a) MotionSense}
\end{minipage}
\hfill
\begin{minipage}[b]{0.43\linewidth}
  \centering
  \centerline{\includegraphics[width=4.7cm]{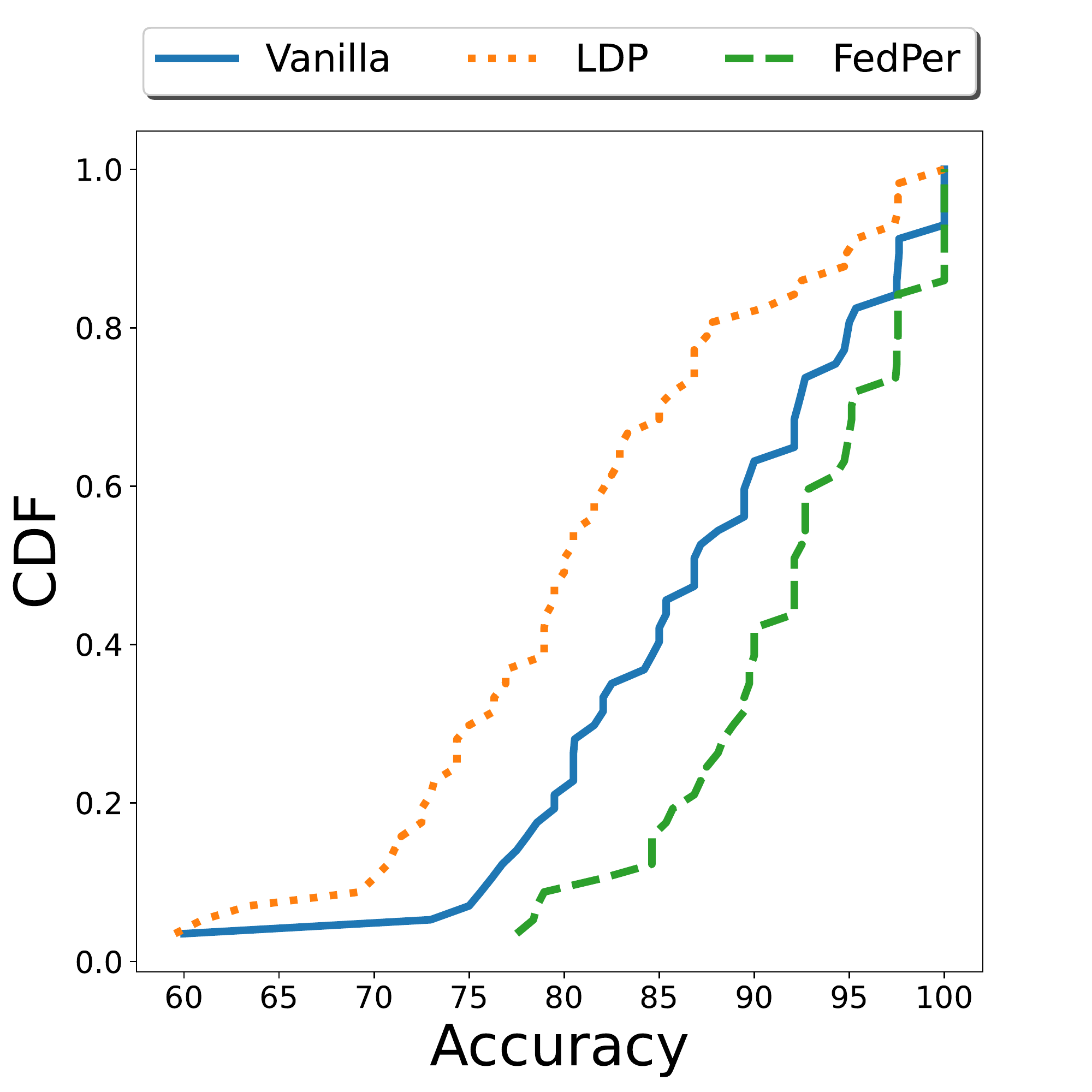}}
  \centerline{(b) MobiAct}
\end{minipage}

\caption{By personalizing upper layers of the model, FedPer sightly increases the accuracy of the activity prediction compared to a FL vanilla approach; local differential privacy, in turn, greatly degrades the accuracy.}
\label{fig:act_motionmobi}
\end{figure}

We measure the accuracy of the activity detection of FedPer and the baselines.
Figure~\ref{fig:act_motionmobi} reports the Cumulative Distribution Function (CDF) of this accuracy over the population of users for MotionSense and MobiAct dataset. 
First, results show that the local adaptation of FedPer on the upper layers sightly increases the accuracy compared to the Vanilla approach (from 1\% to 7\% of increase on average for MotionSense and MobiAct, respectively).
Second, results show that LDP baseline degrades significantly the accuracy for both datasets (10\% on average of MotionSense and 6\% on average for MobiAct).
Indeed, by introducing noise, the convergence of the model is greatly degraded leading to a loss of prediction for all users. This result comforts previous results~\cite{impact1,impact2}.


\begin{figure}[!h]

\begin{minipage}[b]{0.43\linewidth}
  \centering
  \centerline{\includegraphics[width=4.5cm]{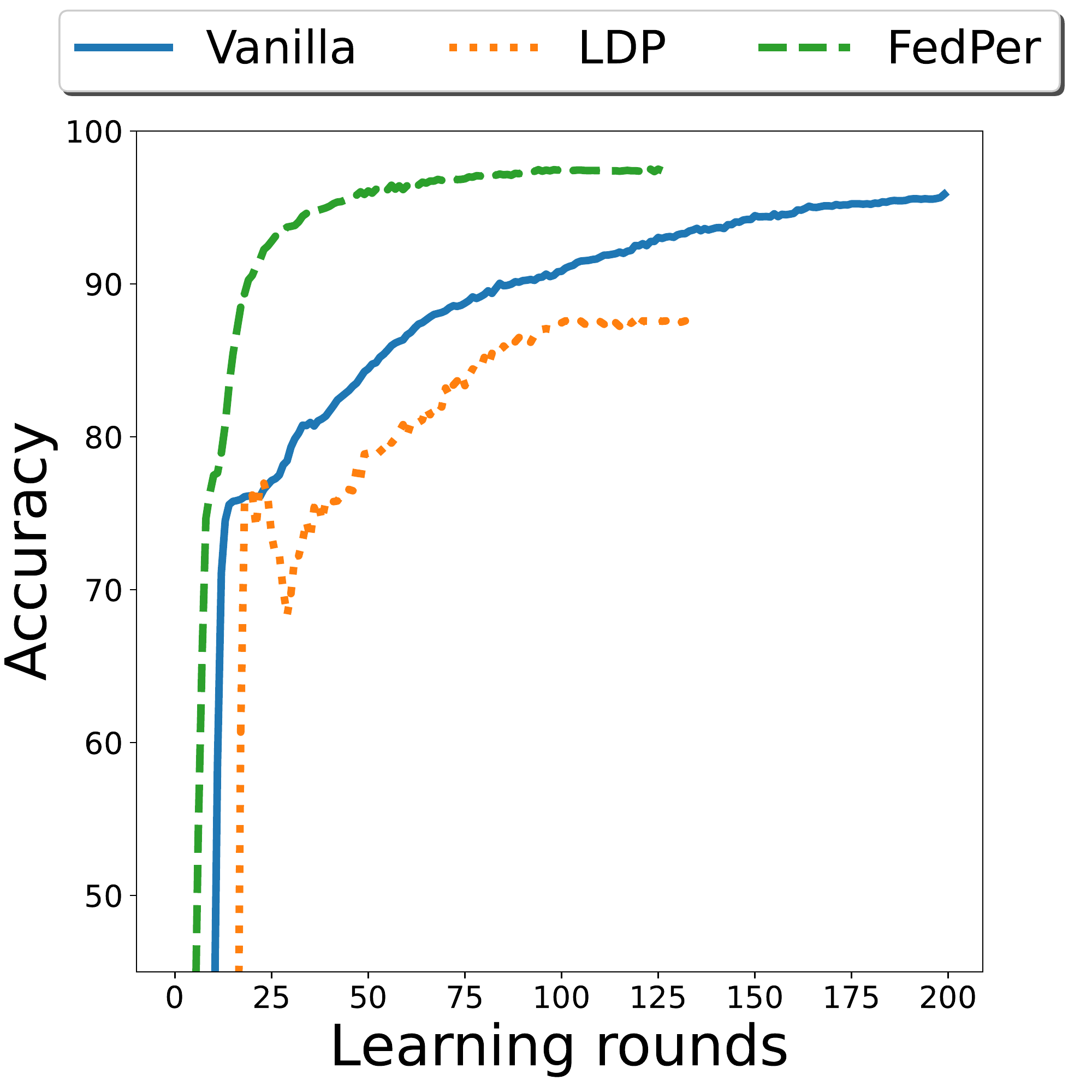}}
  \centerline{(a) MotionSense}
\end{minipage}
\hfill
\begin{minipage}[b]{0.43\linewidth}
  \centering
  \centerline{\includegraphics[width=4.5cm]{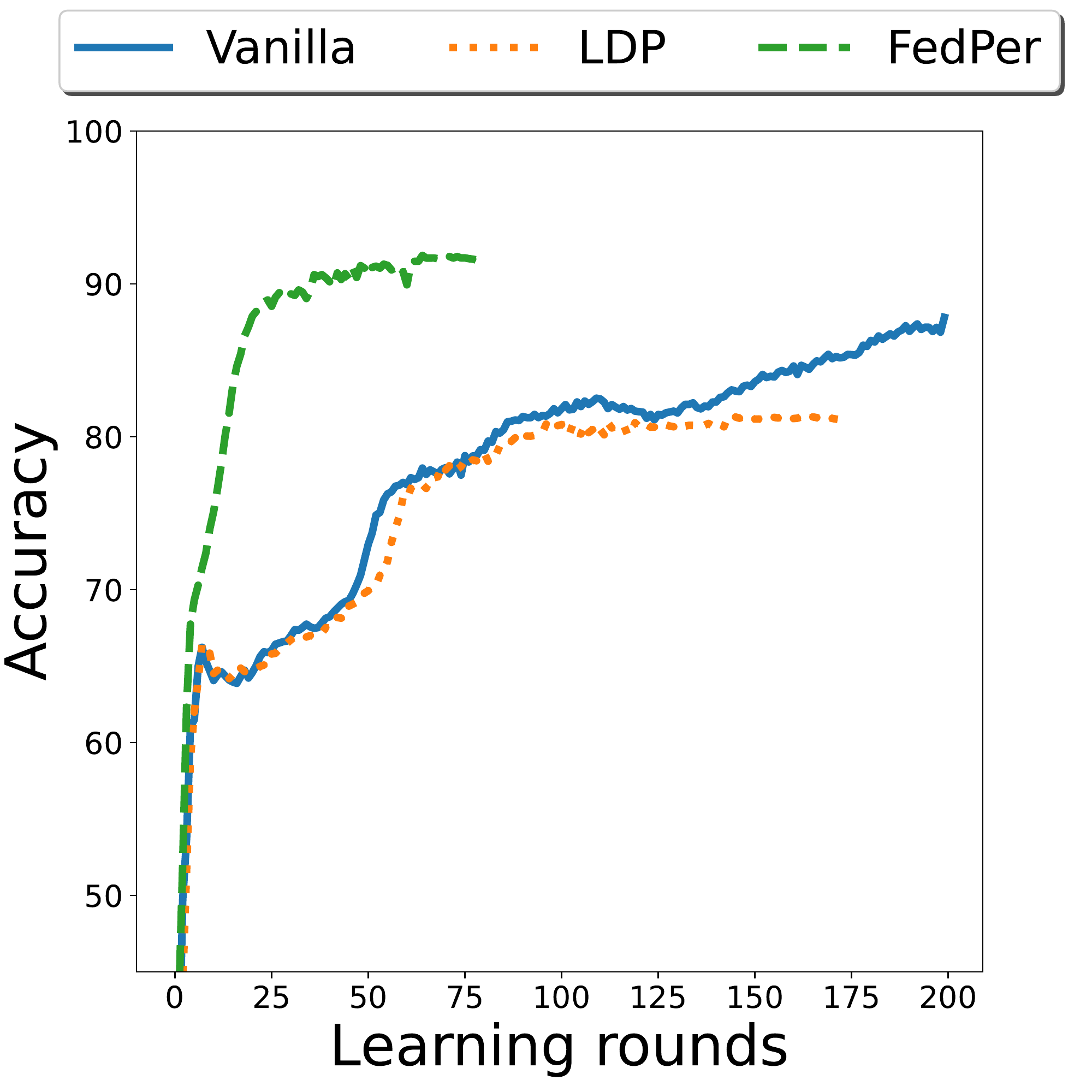}}
  \centerline{(b) MobiAct}
\end{minipage}

\caption{By using personalized layers instead of aggregated information, the learning is drastically speeds up.}
\vspace{-2mm}
\label{fig:act_perep}
\end{figure}

We also measured the convergence speed of the learning.
Figure~\ref{fig:act_perep} depicts the accuracy of the activity detection as a function of learning rounds for FedPer and the Vanilla approach.
Results show that FedPer drastically speeds up the convergence. For instance, FedPer achieves 90\% of accuracy after 12 learning rounds on MotionSense where the Vanilla approach achieves the same accuracy after 100 learning rounds. For MobiAct, FedPer achieves 90\% of accuracy after 35 learning rounds where the Vanilla approach only reaches 86\% of accuracy after 200 learning rounds.
By using its personalized layers at each learning round instead of starting the learning from the aggregate model sent by the server, the accuracy increases faster.
For LDP, we can observe that the noise introduced prevents the model from converging.


\begin{figure}[h!]
\begin{minipage}[b]{0.45\linewidth}
  \centering
  \centerline{\includegraphics[width=4.5cm]{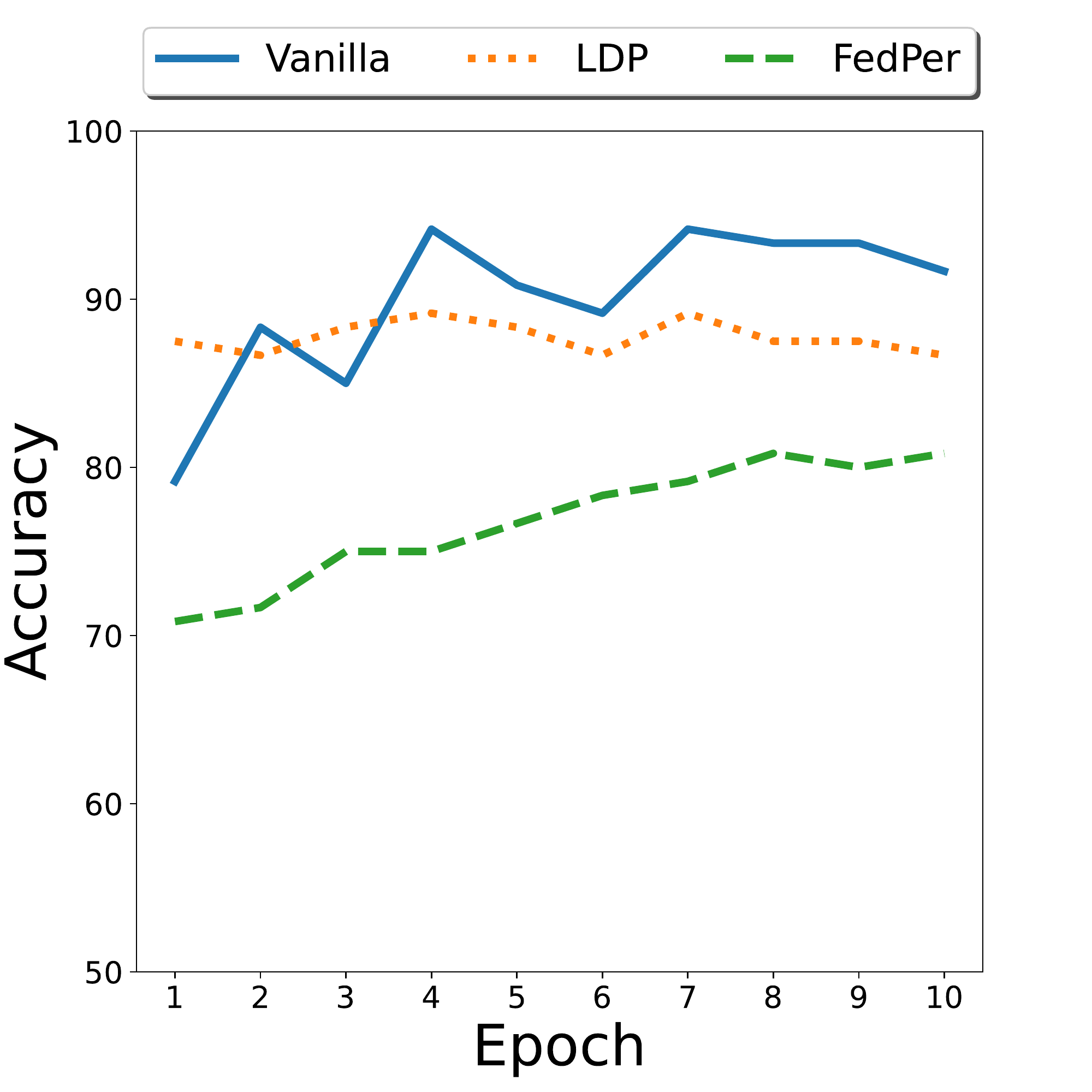}}
  \centerline{(a) Gender - MotionSense}
\end{minipage}
\hfill
\begin{minipage}[b]{0.45\linewidth}
  \centering
  \centerline{\includegraphics[width=4.5cm]{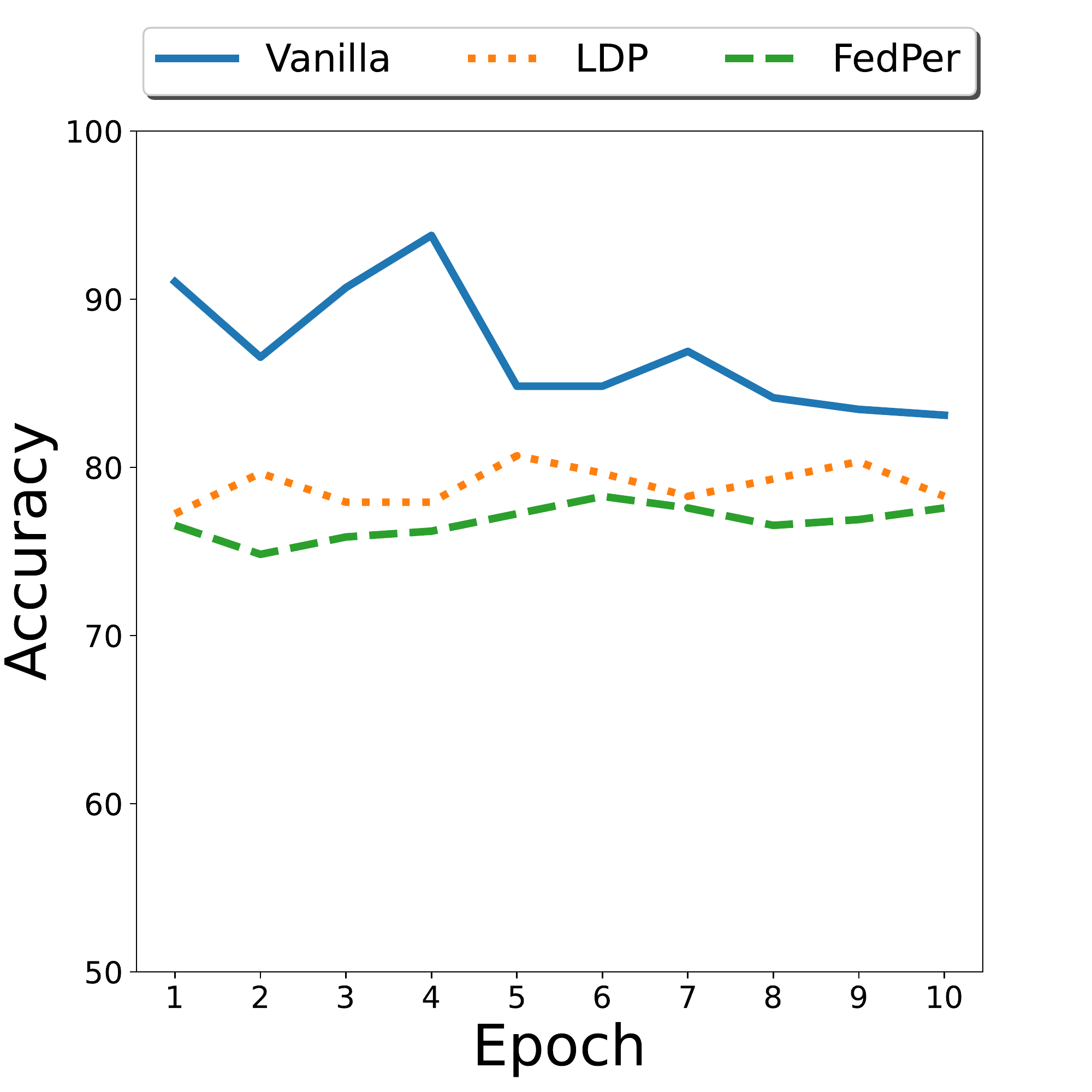}}
  \centerline{(b) Gender - MobiAct}
\end{minipage}
\vfill
\begin{minipage}[b]{0.45\linewidth}
  \centering
  \centerline{\includegraphics[width=4.5cm]{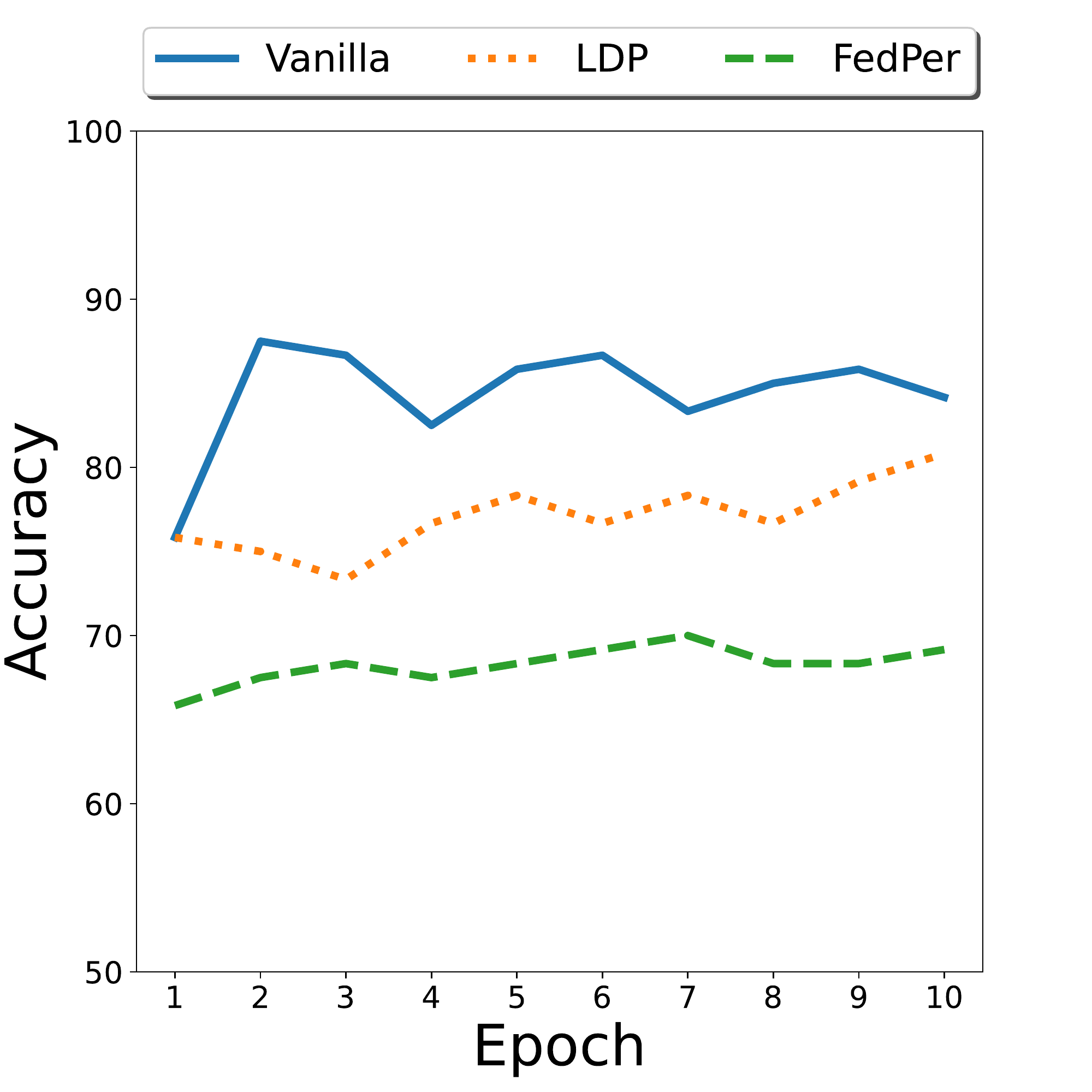}}
  \centerline{(c) BMI - MotionSense}
\end{minipage}
\hfill
\begin{minipage}[b]{0.45\linewidth}
  \centering
  \centerline{\includegraphics[width=4.5cm]{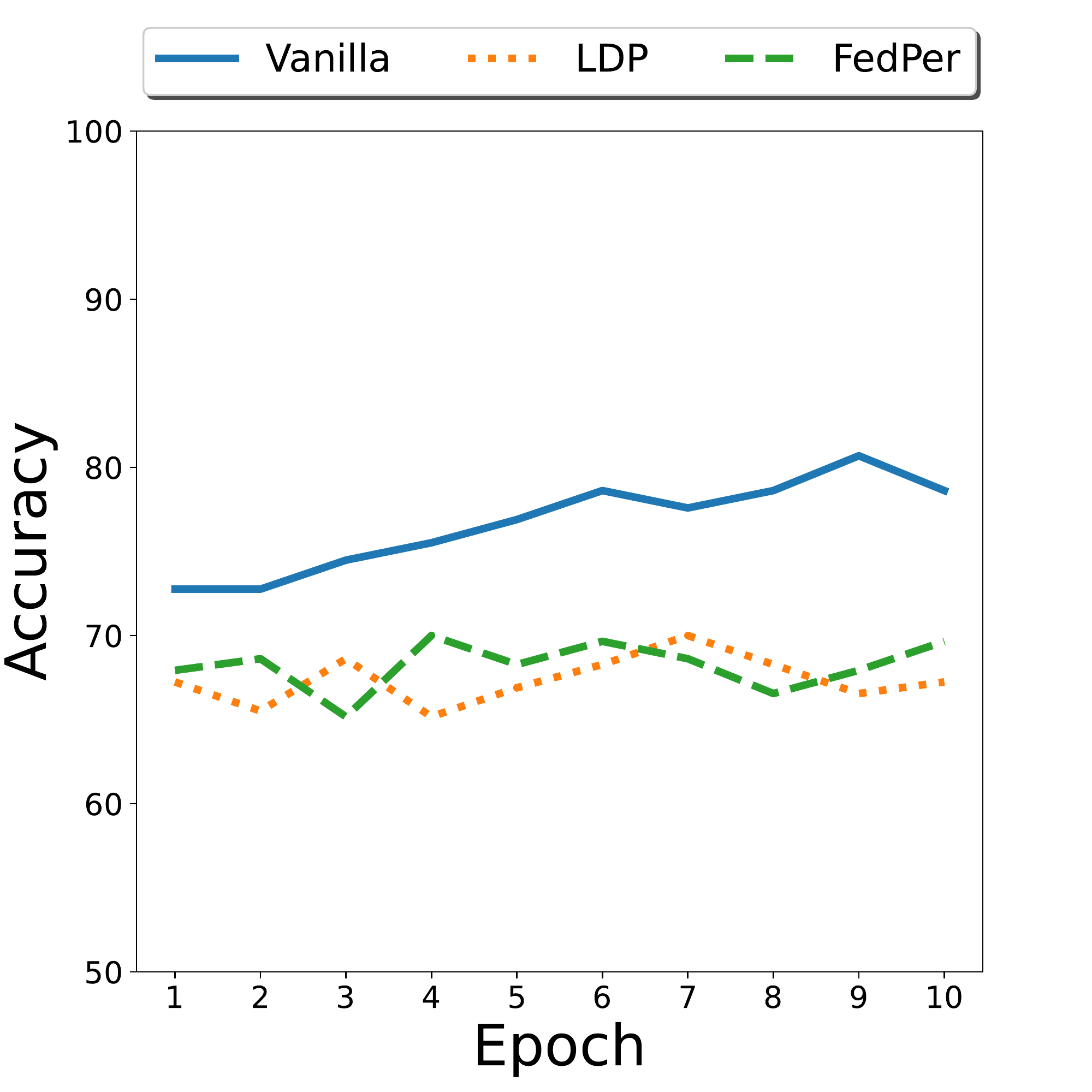}}
  \centerline{(d) BMI - MobiAct}
\end{minipage}
\caption{The increase of number of learning epochs per user increases the accuracy of the attack on both sensitive attributes. }
\vspace{-2mm}
\label{fig:per_ep}
\end{figure}

\subsection{Privacy evaluation through attribute inference}
\label{privacy-attribute}

We conducted an attribute inference attack to infer the gender and the BMI of users from their model updates sent to the server. 
In this attack, participants train their local model on 80\% of their data. Once all the models are sent to the server, only the models from one class of the targeted attribute are aggregated (in our case, models from women for gender inference, and models from overweight users for BMI inference). Then the server sends back the aggregated model to all the users that fine-tune locally on the remaining 20\% of their data (e.g., training from a model aggregating model updates from women) before returning the update to the server. The adversary then trains an RF classifier on these model updates to infer the sensitive attribute. This training exploits 80\% of all the updates and the testing is done on the remaining 20\%, with cross validation.

Figure~\ref{fig:per_ep} evaluates, for both datasets, the accuracy of these both sensitive attribute inferences over the epochs of local learning.
Firstly, results show that without any protection (i.e., the Vanilla approach), all sensitive attributes can be inferred with high accuracy for both datasets (e.g., around 90\% of accuracy for the gender on MotionSense).
FedPer reduces this accuracy between 10\% and 20\% according to the dataset and the sensitive attribute.
Results also show that FedPer better protects users against inference attack compared to LDP regardless of dataset and sensitive attributes (from 5\% to 10\% of accuracy loss for MotionSense).

Secondly, results show that the inference accuracy tends to increase over the epochs for all approaches.
This is explained by the fact that attribute inference attack is closely related to overfitting~\cite{yeom}, the more the model learns on user's data, the more it adjusts the parameters to data structure and the more it may incorporate sensitive information. 



\begin{figure}[!h]

\begin{minipage}[b]{0.45\linewidth}
  \centering
  \centerline{\includegraphics[width=4.5cm]{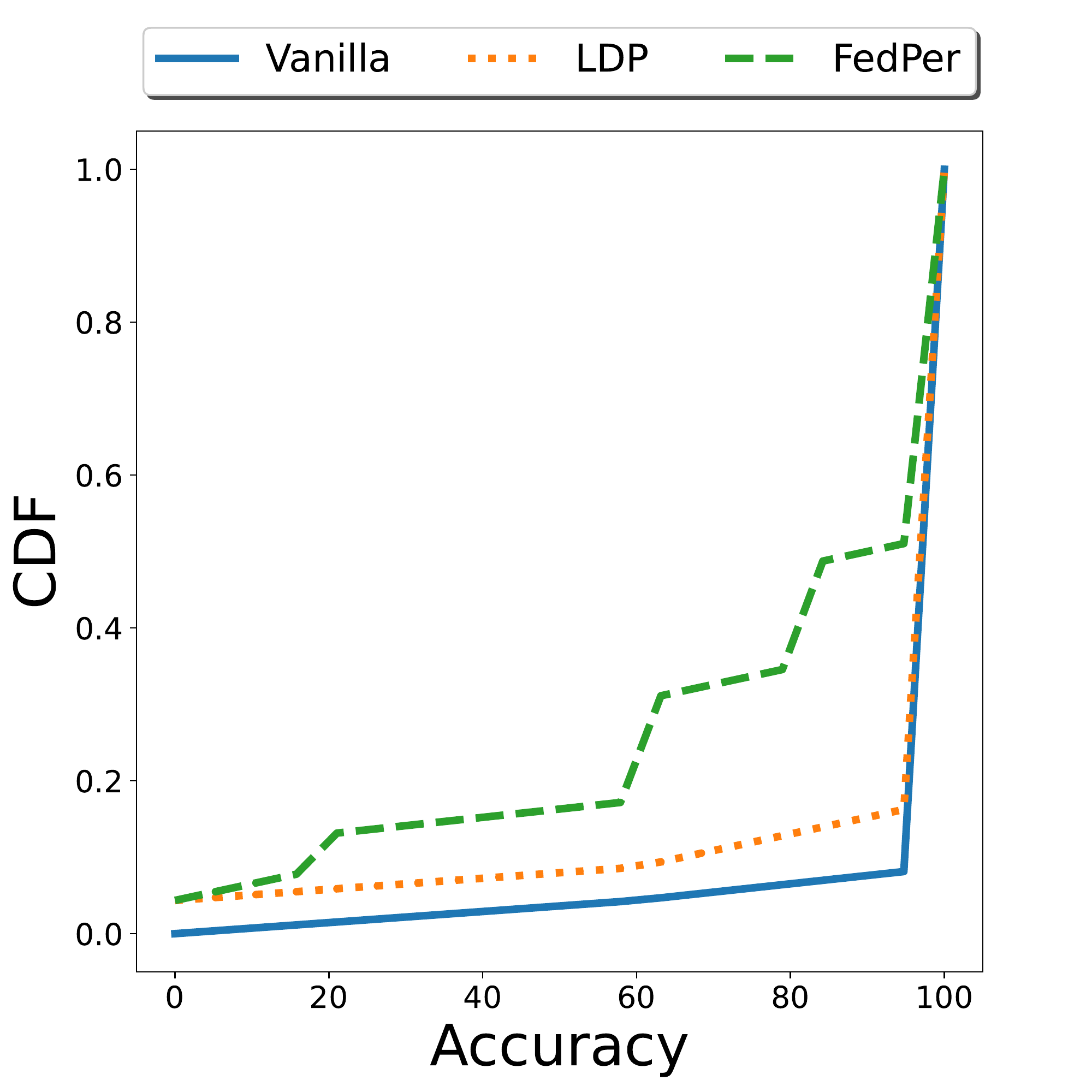}}
  \centerline{(a) Gender - MotionSense}
\end{minipage}
\hfill
\begin{minipage}[b]{0.45\linewidth}
  \centering
  \centerline{\includegraphics[width=4.5cm]{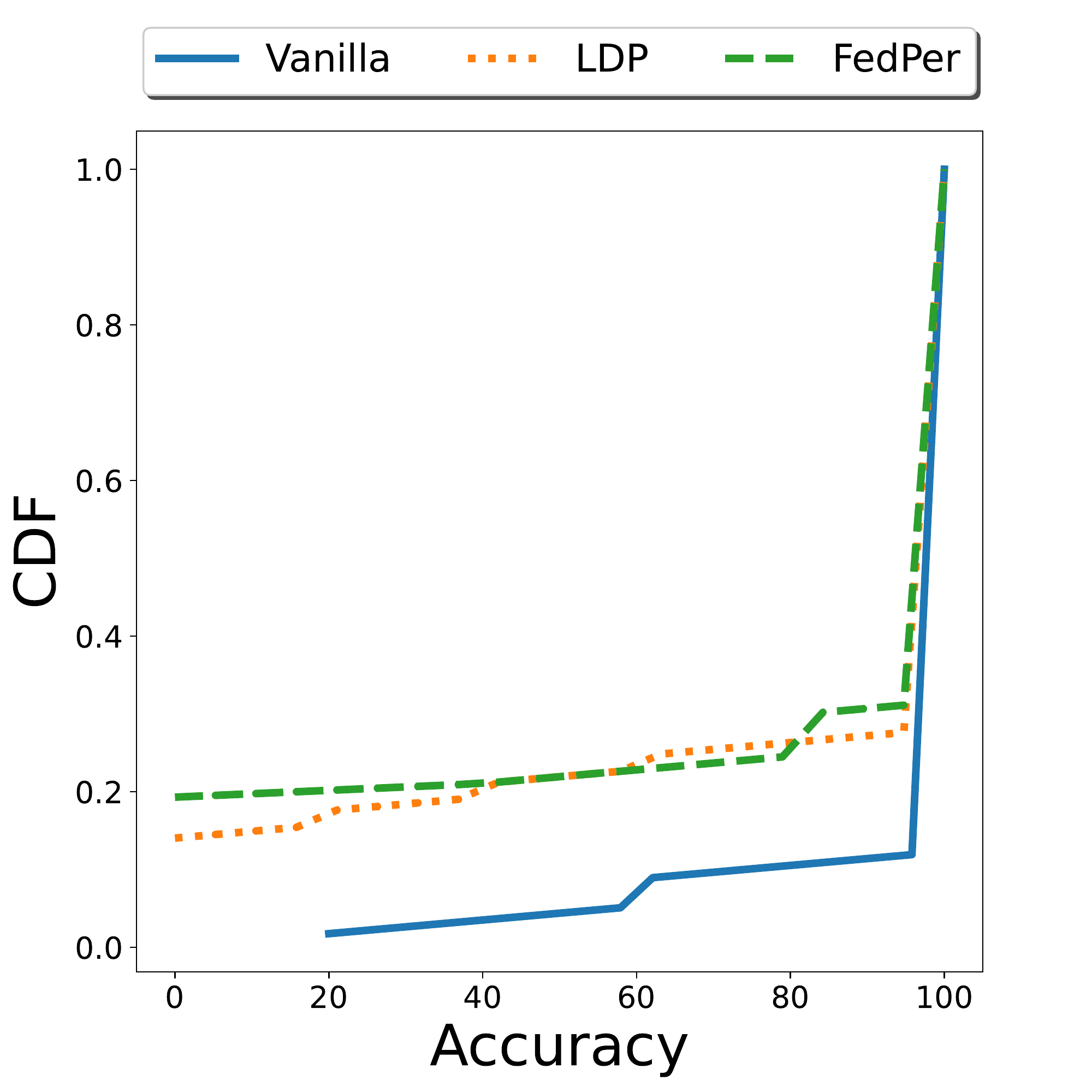}}
  \centerline{(b) Gender - MobiAct}
\end{minipage}
\vfill
\begin{minipage}[b]{0.45\linewidth}
  \centering
  \centerline{\includegraphics[width=4.5cm]{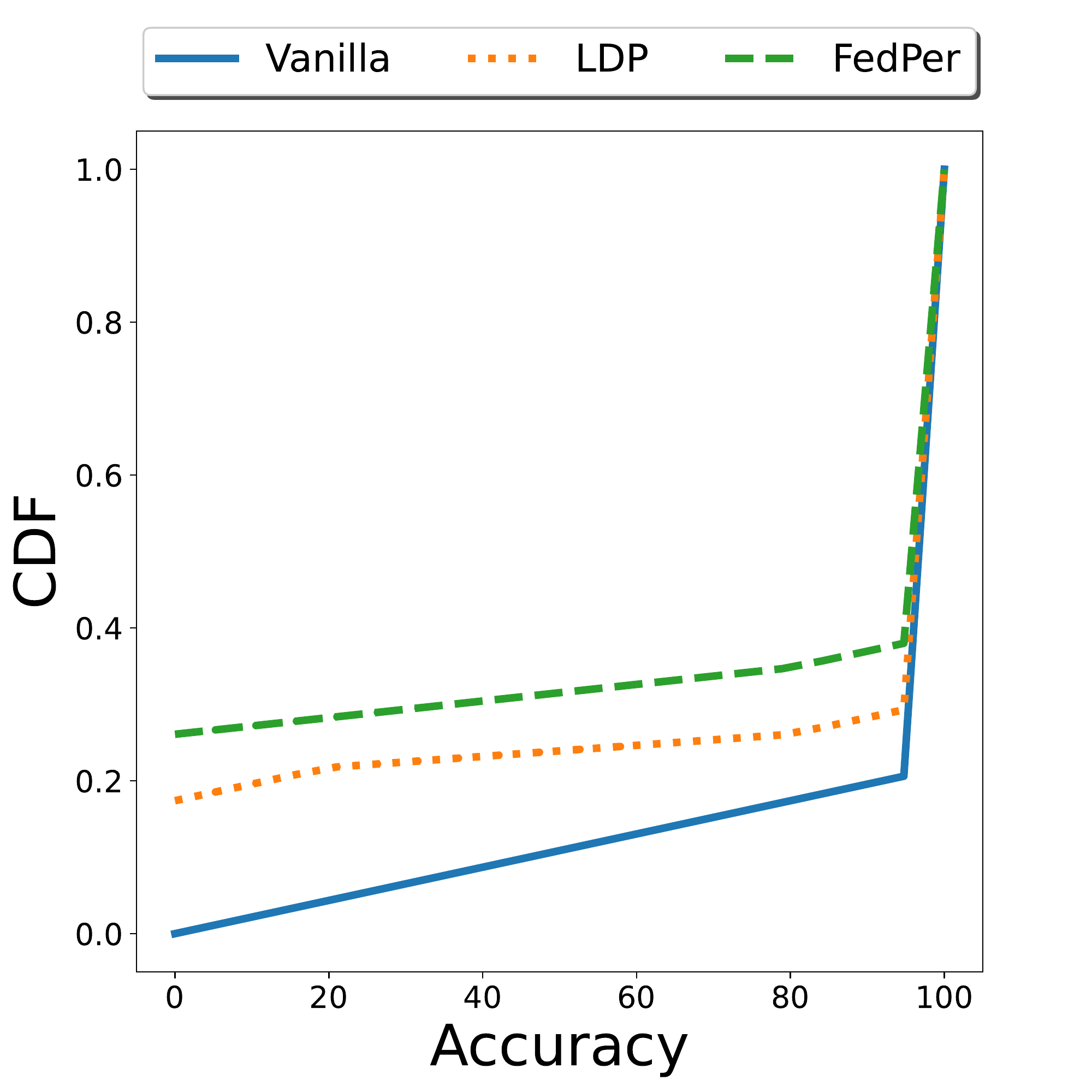}}
  \centerline{(c) BMI - MotionSense}
\end{minipage}
\hfill
\begin{minipage}[b]{0.45\linewidth}
  \centering
  \centerline{\includegraphics[width=4.5cm]{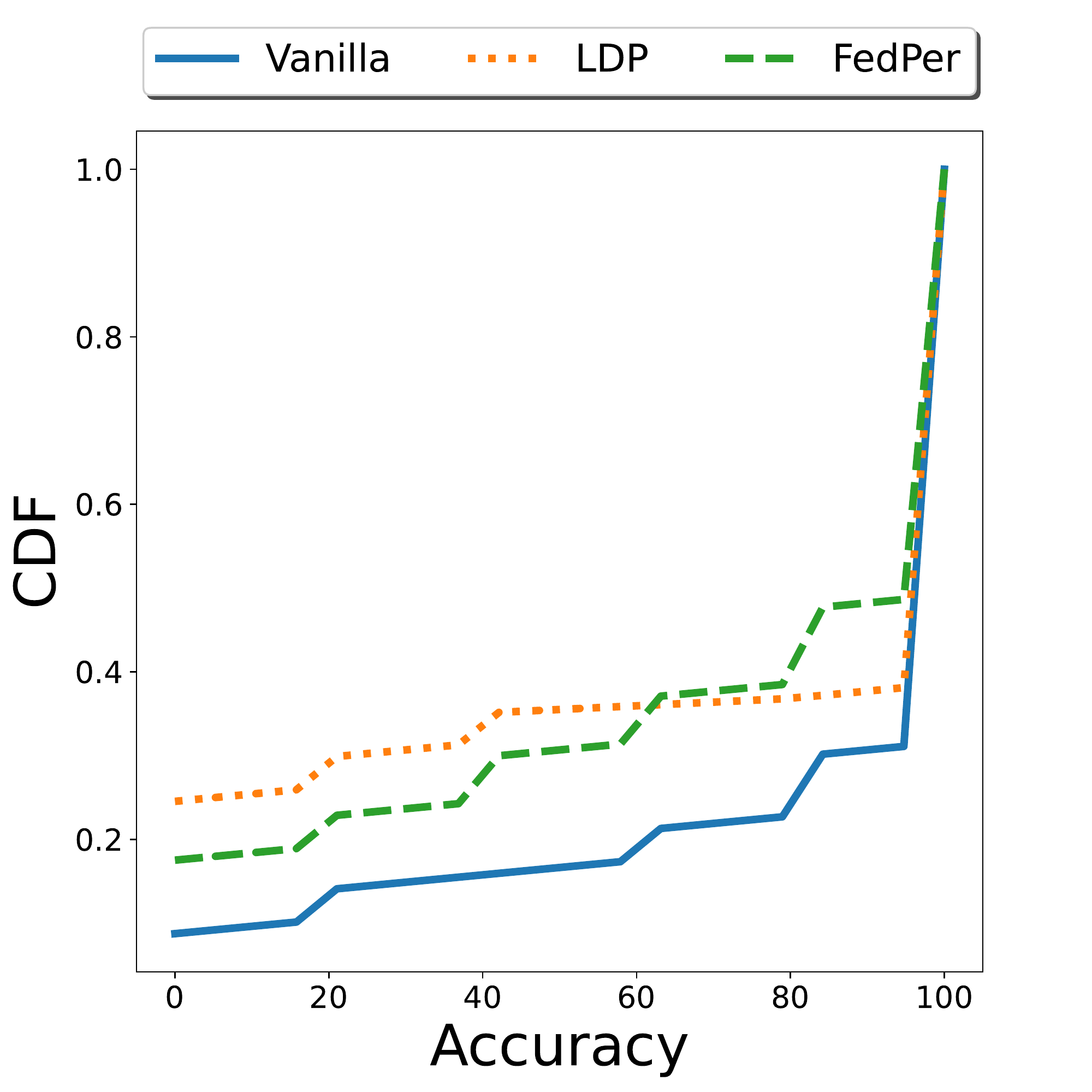}}
  \centerline{(d) BMI - MobiAct}
\end{minipage}

\caption{FedPer and LDP increase the number of users with a small inference accuracy.}
\vspace{-2mm}
\label{fig:selec_ep}
\end{figure}

Figure~\ref{fig:selec_ep} reports the CDF of the inference accuracy over the participants.
Results show that while each attribute can be inferred with high accuracy for a large part of the users, this accuracy drops for few percent of users.
FedPer and LDP increase the percentage of users with a small inference accuracy.


\begin{figure}[!h]
\begin{minipage}[b]{0.48\linewidth}
  \centering
  \centerline{\includegraphics[width=4.5cm]{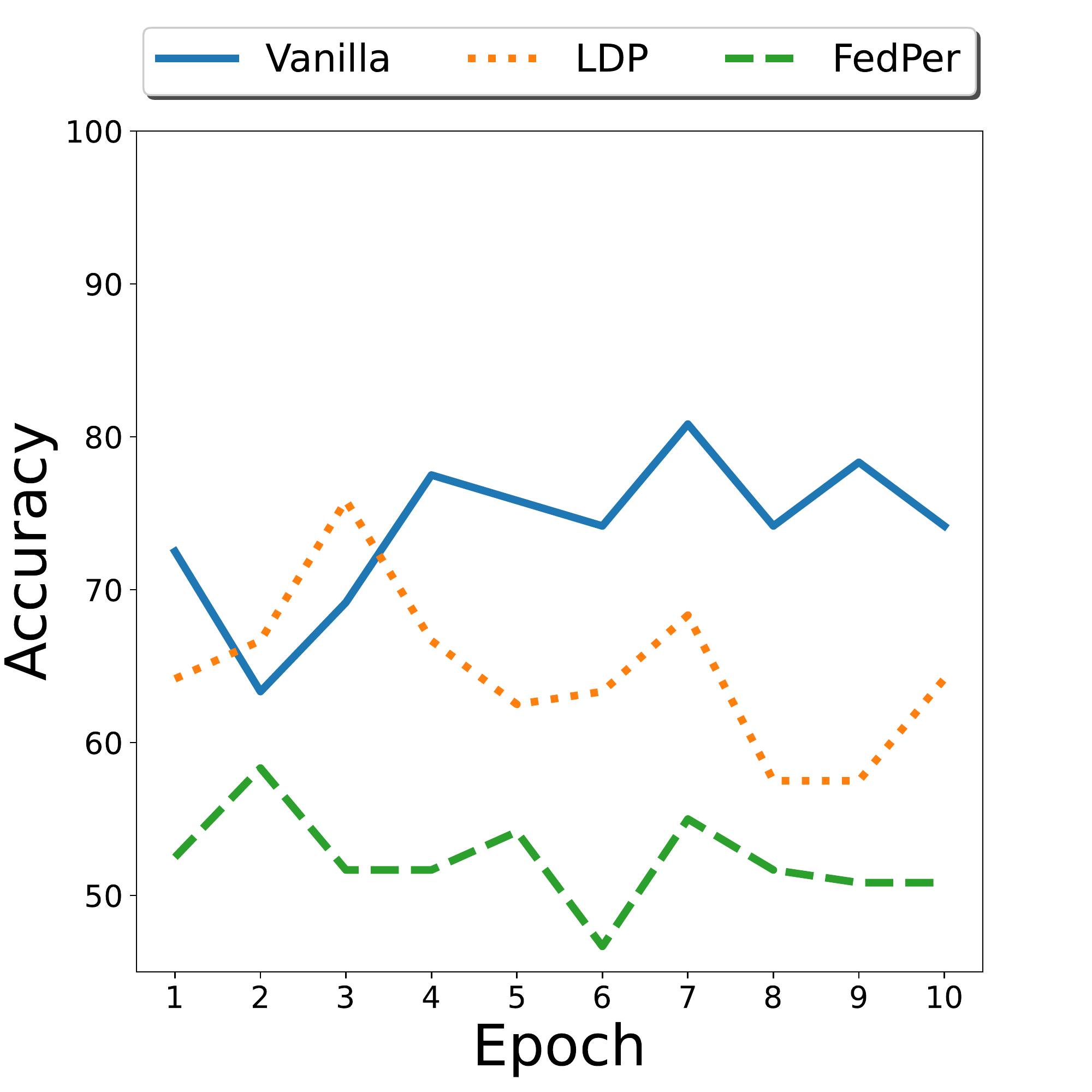}}
  \centerline{(a) MotionSense}
\end{minipage}
\hfill
\begin{minipage}[b]{0.48\linewidth}
  \centering
  \centerline{\includegraphics[width=4.5cm]{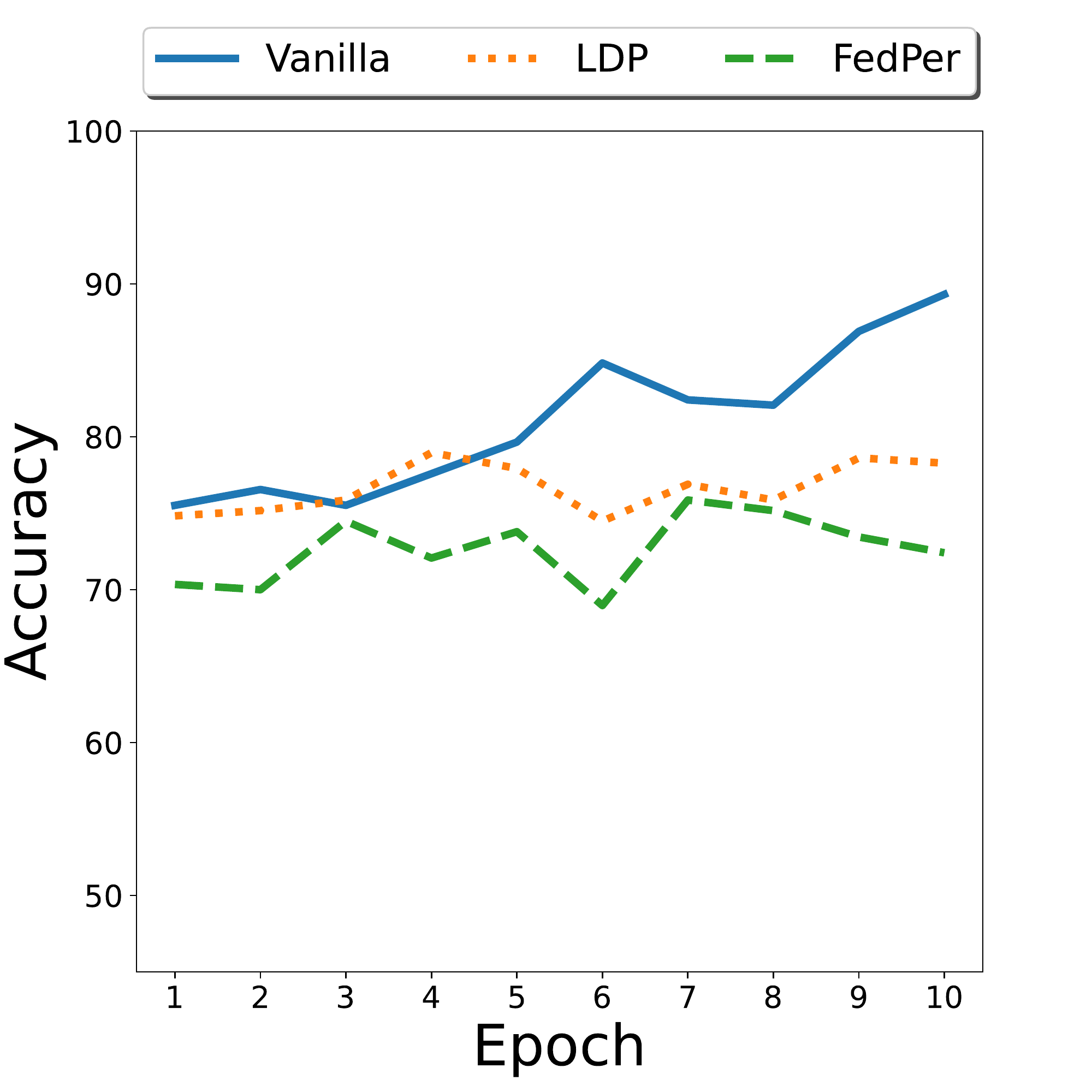}}
  \centerline{(b) MobiAct}
\end{minipage}

\caption{FedPer and LDP significantly decrease the accuracy of the membership inference attack compare to Vanilla method}
\vspace{-3mm}
\label{fig:membership}
\end{figure}

\subsection{Privacy evaluation through membership inference}
\label{privacy-mia}

Lastly, we conduct a membership inference attack to evaluate privacy. In this attack, 50\% of the users follow a normal FL learning round with 80\% of their data. The models are sent to the server which disseminates back the aggregated model to all the users. All of them fine-tune the aggregated model on their remaining 20\% of data. The server then trains a RF to classify membership from model updates for all users (using 80\% of all these updates for the training and 20\% for the testing with cross validation as described in section~\ref{settings}).

Figure~\ref{fig:membership} depicts the accuracy of this inference attack for both datasets and for all approaches.
Similarly to the attribute inference attack, results show that the membership inference attack is more efficient on the Vanilla approach.
FedPer provides the best protection compared to LDP (20\% on average for MotionSense dataset). Interesting enough, FedPer depicts an accuracy close to 50\% which correspond to a random guess (if the data of a specific user has been used to train the model) for MotionSense dataset.



\section{Conclusion}
\label{ending}

We experimentally quantified the utility and privacy trade-off of FL using private personalized layers proposed by~\cite{fedper} in a context of activity recognition. 
We consider both an attribute and a membership inference attack to measure privacy leakage.
Results show that using private personalized layers provides a better utility and privacy trade-off compared to a FL vanilla approach and a defense scheme using local differential privacy.
These results tend to show that minimizing the information exchanged with the server is an interesting avenue for both improving accuracy and limiting privacy leakage.
To comfort these results, it would be interesting to use other quantitative privacy metrics such as \textit{average information leakage} and \textit{maximum information leakage}~\cite{othermetrics}.

\bibliographystyle{ACM-Reference-Format}
\bibliography{example_paper}


\begin{thebibliography}{22}


\ifx \showCODEN    \undefined \def \showCODEN     #1{\unskip}     \fi
\ifx \showDOI      \undefined \def \showDOI       #1{#1}\fi
\ifx \showISBNx    \undefined \def \showISBNx     #1{\unskip}     \fi
\ifx \showISBNxiii \undefined \def \showISBNxiii  #1{\unskip}     \fi
\ifx \showISSN     \undefined \def \showISSN      #1{\unskip}     \fi
\ifx \showLCCN     \undefined \def \showLCCN      #1{\unskip}     \fi
\ifx \shownote     \undefined \def \shownote      #1{#1}          \fi
\ifx \showarticletitle \undefined \def \showarticletitle #1{#1}   \fi
\ifx \showURL      \undefined \def \showURL       {\relax}        \fi
\providecommand\bibfield[2]{#2}
\providecommand\bibinfo[2]{#2}
\providecommand\natexlab[1]{#1}
\providecommand\showeprint[2][]{arXiv:#2}

\bibitem[\protect\citeauthoryear{Arivazhagan, Aggarwal, Singh, and
  Choudhary}{Arivazhagan et~al\mbox{.}}{2019}]%
        {fedper}
\bibfield{author}{\bibinfo{person}{M.~G. Arivazhagan}, \bibinfo{person}{V.
  Aggarwal}, \bibinfo{person}{A.~K. Singh}, {and} \bibinfo{person}{S.
  Choudhary}.} \bibinfo{year}{2019}\natexlab{}.
\newblock \showarticletitle{Federated Learning with Personalization Layers}.
\newblock \bibinfo{journal}{\emph{CoRR}}  \bibinfo{volume}{abs/1912.00818}
  (\bibinfo{year}{2019}).
\newblock


\bibitem[\protect\citeauthoryear{Bagdasaryan, Veit, Hua, Estrin, and
  Shmatikov}{Bagdasaryan et~al\mbox{.}}{2018}]%
        {backdoor}
\bibfield{author}{\bibinfo{person}{E. Bagdasaryan}, \bibinfo{person}{A. Veit},
  \bibinfo{person}{Y. Hua}, \bibinfo{person}{D. Estrin}, {and}
  \bibinfo{person}{V. Shmatikov}.} \bibinfo{year}{2018}\natexlab{}.
\newblock \showarticletitle{How To Backdoor Federated Learning}.
\newblock \bibinfo{journal}{\emph{CoRR}}  \bibinfo{volume}{abs/1807.00459}
  (\bibinfo{year}{2018}).
\newblock


\bibitem[\protect\citeauthoryear{Bernstein, Zhao, Azizzadenesheli, and
  Anandkumar}{Bernstein et~al\mbox{.}}{2018}]%
        {bernstein}
\bibfield{author}{\bibinfo{person}{J. Bernstein}, \bibinfo{person}{J. Zhao},
  \bibinfo{person}{K. Azizzadenesheli}, {and} \bibinfo{person}{A. Anandkumar}.}
  \bibinfo{year}{2018}\natexlab{}.
\newblock \showarticletitle{signSGD with Majority Vote is Communication
  Efficient And Byzantine Fault Tolerant}.
\newblock \bibinfo{journal}{\emph{CoRR}}  \bibinfo{volume}{abs/1802.04434}
  (\bibinfo{year}{2018}).
\newblock


\bibitem[\protect\citeauthoryear{Blanchard, El~Mhamdi, Guerraoui, and
  Stainer}{Blanchard et~al\mbox{.}}{2017}]%
        {blanchard}
\bibfield{author}{\bibinfo{person}{P. Blanchard}, \bibinfo{person}{E.~M.
  El~Mhamdi}, \bibinfo{person}{R. Guerraoui}, {and} \bibinfo{person}{J.
  Stainer}.} \bibinfo{year}{2017}\natexlab{}.
\newblock \showarticletitle{Machine Learning with Adversaries: Byzantine
  Tolerant Gradient Descent}.
\newblock \bibinfo{journal}{\emph{NIPS'17}}  \bibinfo{volume}{30}
  (\bibinfo{year}{2017}), \bibinfo{pages}{118–128}.
\newblock


\bibitem[\protect\citeauthoryear{Boutet, Frindel, Gambs, Jourdan, and
  Ngueveu}{Boutet et~al\mbox{.}}{2021}]%
        {dysan}
\bibfield{author}{\bibinfo{person}{A. Boutet}, \bibinfo{person}{C. Frindel},
  \bibinfo{person}{S. Gambs}, \bibinfo{person}{T. Jourdan}, {and}
  \bibinfo{person}{C.~R. Ngueveu}.} \bibinfo{year}{2021}\natexlab{}.
\newblock \showarticletitle{DYSAN: Dynamically sanitizing motion sensor data
  against sensitive inferences through adversarial networks}.
\newblock \bibinfo{journal}{\emph{AsiaCCS'21}} (\bibinfo{year}{2021}).
\newblock


\bibitem[\protect\citeauthoryear{du~Pin~Calmon and Fawaz}{du~Pin~Calmon and
  Fawaz}{2012}]%
        {othermetrics}
\bibfield{author}{\bibinfo{person}{F. du Pin~Calmon} {and} \bibinfo{person}{N.
  Fawaz}.} \bibinfo{year}{2012}\natexlab{}.
\newblock \showarticletitle{Privacy Against Statistical Inference}.
\newblock \bibinfo{journal}{\emph{CoRR}}  \bibinfo{volume}{abs/1210.2123}
  (\bibinfo{year}{2012}).
\newblock


\bibitem[\protect\citeauthoryear{Fang and Qian}{Fang and Qian}{2021}]%
        {hc_smc}
\bibfield{author}{\bibinfo{person}{H. Fang} {and} \bibinfo{person}{Q. Qian}.}
  \bibinfo{year}{2021}\natexlab{}.
\newblock \showarticletitle{Privacy Preserving Machine Learning with
  Homomorphic Encryption and Federated Learning}.
\newblock \bibinfo{journal}{\emph{Future Internet}} \bibinfo{volume}{13},
  \bibinfo{number}{4} (\bibinfo{year}{2021}), \bibinfo{pages}{94}.
\newblock


\bibitem[\protect\citeauthoryear{Ganju, Wang, Yang, Gunter, and Borisov}{Ganju
  et~al\mbox{.}}{2018}]%
        {attribute}
\bibfield{author}{\bibinfo{person}{K. Ganju}, \bibinfo{person}{Q. Wang},
  \bibinfo{person}{W. Yang}, \bibinfo{person}{C.~A. Gunter}, {and}
  \bibinfo{person}{N. Borisov}.} \bibinfo{year}{2018}\natexlab{}.
\newblock \showarticletitle{Property Inference Attacks on Fully Connected
  Neural Networks Using Permutation Invariant Representations}.
\newblock \bibinfo{journal}{\emph{ACM SIGSAC'18}} (\bibinfo{year}{2018}),
  \bibinfo{pages}{619–633}.
\newblock


\bibitem[\protect\citeauthoryear{Kr\"{o}ger, Raschke, and Bhuiyan}{Kr\"{o}ger
  et~al\mbox{.}}{2019}]%
        {privacyhealth}
\bibfield{author}{\bibinfo{person}{J.~L. Kr\"{o}ger}, \bibinfo{person}{P.
  Raschke}, {and} \bibinfo{person}{T.~R. Bhuiyan}.}
  \bibinfo{year}{2019}\natexlab{}.
\newblock \showarticletitle{Privacy Implications of Accelerometer Data: A
  Review of Possible Inferences}.
\newblock \bibinfo{journal}{\emph{ICCSP'19}} (\bibinfo{year}{2019}),
  \bibinfo{pages}{81–87}.
\newblock


\bibitem[\protect\citeauthoryear{Li, Sahu, Talwalkar, and Smith}{Li
  et~al\mbox{.}}{2020}]%
        {review_fl}
\bibfield{author}{\bibinfo{person}{T. Li}, \bibinfo{person}{A.~K. Sahu},
  \bibinfo{person}{A. Talwalkar}, {and} \bibinfo{person}{V. Smith}.}
  \bibinfo{year}{2020}\natexlab{}.
\newblock \showarticletitle{Federated Learning: Challenges, Methods, and Future
  Directions}.
\newblock \bibinfo{journal}{\emph{IEEE Signal Process. Mag.}}
  \bibinfo{volume}{37}, \bibinfo{number}{3} (\bibinfo{year}{2020}),
  \bibinfo{pages}{50--60}.
\newblock


\bibitem[\protect\citeauthoryear{Malekzadeh, Clegg, Cavallaro, and
  Haddadi}{Malekzadeh et~al\mbox{.}}{2019}]%
        {motionsense}
\bibfield{author}{\bibinfo{person}{M. Malekzadeh}, \bibinfo{person}{R.~G.
  Clegg}, \bibinfo{person}{A. Cavallaro}, {and} \bibinfo{person}{H. Haddadi}.}
  \bibinfo{year}{2019}\natexlab{}.
\newblock \showarticletitle{Mobile Sensor Data Anonymization}.
\newblock \bibinfo{journal}{\emph{ACM IoTDI'19}} (\bibinfo{year}{2019}),
  \bibinfo{pages}{49--58}.
\newblock


\bibitem[\protect\citeauthoryear{McMahan, Moore, Ramage, and y~Arcas}{McMahan
  et~al\mbox{.}}{2016}]%
        {mcmahan}
\bibfield{author}{\bibinfo{person}{H.~B. McMahan}, \bibinfo{person}{E. Moore},
  \bibinfo{person}{D. Ramage}, {and} \bibinfo{person}{B.~A. y Arcas}.}
  \bibinfo{year}{2016}\natexlab{}.
\newblock \showarticletitle{Federated Learning of Deep Networks using Model
  Averaging}.
\newblock \bibinfo{journal}{\emph{CoRR}}  \bibinfo{volume}{abs/1602.05629}
  (\bibinfo{year}{2016}).
\newblock


\bibitem[\protect\citeauthoryear{Naseri, Hayes, and Cristofaro}{Naseri
  et~al\mbox{.}}{2020}]%
        {naseri}
\bibfield{author}{\bibinfo{person}{M. Naseri}, \bibinfo{person}{J. Hayes},
  {and} \bibinfo{person}{E.~De Cristofaro}.} \bibinfo{year}{2020}\natexlab{}.
\newblock \showarticletitle{Toward Robustness and Privacy in Federated
  Learning: Experimenting with Local and Central Differential Privacy}.
\newblock \bibinfo{journal}{\emph{CoRR}}  \bibinfo{volume}{abs/2009.03561}
  (\bibinfo{year}{2020}).
\newblock


\bibitem[\protect\citeauthoryear{Nasr, Shokri, and Houmansadr}{Nasr
  et~al\mbox{.}}{2019}]%
        {passiveactive2}
\bibfield{author}{\bibinfo{person}{M. Nasr}, \bibinfo{person}{R. Shokri}, {and}
  \bibinfo{person}{A. Houmansadr}.} \bibinfo{year}{2019}\natexlab{}.
\newblock \showarticletitle{Comprehensive Privacy Analysis of Deep Learning:
  Passive and Active White-box Inference Attacks against Centralized and
  Federated Learning}.
\newblock \bibinfo{journal}{\emph{IEEE SP'19}} (\bibinfo{year}{2019}),
  \bibinfo{pages}{739--753}.
\newblock


\bibitem[\protect\citeauthoryear{Park, Chang, and Nam}{Park
  et~al\mbox{.}}{2017}]%
        {health2}
\bibfield{author}{\bibinfo{person}{E. Park}, \bibinfo{person}{H.-J. Chang},
  {and} \bibinfo{person}{H.~S. Nam}.} \bibinfo{year}{2017}\natexlab{}.
\newblock \showarticletitle{Use of Machine Learning Classifiers and Sensor Data
  to Detect Neurological Deficit in Stroke Patients}.
\newblock \bibinfo{journal}{\emph{J Med Internet Res}} \bibinfo{volume}{19},
  \bibinfo{number}{4} (\bibinfo{year}{2017}), \bibinfo{pages}{e120}.
\newblock


\bibitem[\protect\citeauthoryear{Qi, Yang, Fan, and Deng}{Qi
  et~al\mbox{.}}{2015}]%
        {health1}
\bibfield{author}{\bibinfo{person}{J. Qi}, \bibinfo{person}{P. Yang},
  \bibinfo{person}{D. Fan}, {and} \bibinfo{person}{Z. Deng}.}
  \bibinfo{year}{2015}\natexlab{}.
\newblock \showarticletitle{A Survey of Physical Activity Monitoring and
  Assessment Using Internet of Things Technology}.
\newblock \bibinfo{journal}{\emph{IEEE CIT'15}} (\bibinfo{year}{2015}),
  \bibinfo{pages}{2353--2358}.
\newblock


\bibitem[\protect\citeauthoryear{Shokri, Stronati, Song, and Shmatikov}{Shokri
  et~al\mbox{.}}{2017}]%
        {membership}
\bibfield{author}{\bibinfo{person}{R. Shokri}, \bibinfo{person}{M. Stronati},
  \bibinfo{person}{C. Song}, {and} \bibinfo{person}{V. Shmatikov}.}
  \bibinfo{year}{2017}\natexlab{}.
\newblock \showarticletitle{Membership Inference Attacks Against Machine
  Learning Models}.
\newblock \bibinfo{journal}{\emph{IEEE SP'17}} (\bibinfo{year}{2017}),
  \bibinfo{pages}{3--18}.
\newblock


\bibitem[\protect\citeauthoryear{van~der Maaten and Hannun}{van~der Maaten and
  Hannun}{2020}]%
        {impact1}
\bibfield{author}{\bibinfo{person}{L. van~der Maaten} {and}
  \bibinfo{person}{A.~Y. Hannun}.} \bibinfo{year}{2020}\natexlab{}.
\newblock \showarticletitle{The Trade-Offs of Private Prediction}.
\newblock \bibinfo{journal}{\emph{CoRR}}  \bibinfo{volume}{abs/2007.05089}
  (\bibinfo{year}{2020}).
\newblock


\bibitem[\protect\citeauthoryear{Vavoulas, Chatzaki, Malliotakis, Pediaditis,
  and Tsiknakis}{Vavoulas et~al\mbox{.}}{2016}]%
        {mobiact}
\bibfield{author}{\bibinfo{person}{G. Vavoulas}, \bibinfo{person}{C. Chatzaki},
  \bibinfo{person}{T. Malliotakis}, \bibinfo{person}{M. Pediaditis}, {and}
  \bibinfo{person}{M. Tsiknakis}.} \bibinfo{year}{2016}\natexlab{}.
\newblock \showarticletitle{The MobiAct Dataset: Recognition of Activities of
  Daily Living using Smartphones}.
\newblock \bibinfo{journal}{\emph{ICT4AWE'16}} (\bibinfo{year}{2016}),
  \bibinfo{pages}{143--151}.
\newblock


\bibitem[\protect\citeauthoryear{Xu and Li}{Xu and Li}{2020}]%
        {passiveactive1}
\bibfield{author}{\bibinfo{person}{M. Xu} {and} \bibinfo{person}{X. Li}.}
  \bibinfo{year}{2020}\natexlab{}.
\newblock \showarticletitle{Subject Property Inference Attack in Collaborative
  Learning}.
\newblock \bibinfo{journal}{\emph{IHMSC'20}} (\bibinfo{year}{2020}),
  \bibinfo{pages}{227--231}.
\newblock


\bibitem[\protect\citeauthoryear{Yeom, Fredrikson, and Jha}{Yeom
  et~al\mbox{.}}{2017}]%
        {yeom}
\bibfield{author}{\bibinfo{person}{S. Yeom}, \bibinfo{person}{M. Fredrikson},
  {and} \bibinfo{person}{S. Jha}.} \bibinfo{year}{2017}\natexlab{}.
\newblock \showarticletitle{The Unintended Consequences of Overfitting:
  Training Data Inference Attacks}.
\newblock \bibinfo{journal}{\emph{CoRR}}  \bibinfo{volume}{abs/1709.01604}
  (\bibinfo{year}{2017}).
\newblock


\bibitem[\protect\citeauthoryear{Yu, Bagdasaryan, and Shmatikov}{Yu
  et~al\mbox{.}}{2020}]%
        {impact2}
\bibfield{author}{\bibinfo{person}{T. Yu}, \bibinfo{person}{E. Bagdasaryan},
  {and} \bibinfo{person}{V. Shmatikov}.} \bibinfo{year}{2020}\natexlab{}.
\newblock \showarticletitle{Salvaging Federated Learning by Local Adaptation}.
\newblock \bibinfo{journal}{\emph{CoRR}}  \bibinfo{volume}{abs/2002.04758}
  (\bibinfo{year}{2020}).
\newblock


\end{thebibliography}

\end{document}